\documentclass[11pt,a4paper]{article}

\usepackage[T1]{fontenc}
\usepackage[utf8]{inputenc}
\usepackage{lmodern}
\usepackage{microtype}
\usepackage[margin=1in]{geometry}
\usepackage{booktabs}
\usepackage{tabularx}
\usepackage{caption}
\usepackage{enumitem}
\usepackage{setspace}
\usepackage[hidelinks,breaklinks=true,bookmarksnumbered=true,%
            bookmarksopen=true,bookmarksopenlevel=1]{hyperref}
\usepackage{xurl}

\setlist{itemsep=0.2em,parsep=0.2em}
\title{\textbf{Inferential Capability Does Not Determine Legal Scope}\\[0.4em]
\large Agentic AI and the Two Concepts of Inference in EU Law}

\author{Nicola Fabiano\thanks{Studio Legale Fabiano, Italy. Independent Researcher on Artificial Intelligence, Data Protection, and Privacy. Expert in the EDPB's Support Pool of Experts --- Field B: Legal Expertise in New Technologies (project-based). Member, International Neural Network Society (INNS). Member, United Nations University AI Network (UNU AI Network). Member, IEEE - IEEE SA. Member, Editorial Advisory Board, Journal of Systemics, Cybernetics and Informatics (JSCI). Member, International Institute of Informatics and Systemics (IIIS). Email: \texttt{info@fabiano.law}. ORCID: 0000-0002-8188-7656.}}

\date{10 August 2026}

\begin{document}

\maketitle
\vspace{-1em}

\begin{abstract}
\small
\setlength{\parskip}{0.2em}
\noindent Two instruments of EU digital law place inference at their centre and mean different things by it. Article 3(1) of the AI Act uses the capability to infer \emph{constitutively}: it is the central feature separating the regulated category from conventional software. The GDPR never defines inference, yet governs it \emph{protectively}: the consequences follow from the processing of personal data and from what the inference says about, or does to, a person, whether or not the technology that produced it qualifies as an AI system.

The two perimeters are not concentric. Their non-coincidence remained invisible in single-shot systems; agentic architectures make it operationally acute. The thesis: \emph{inferential capability does not determine legal scope, and its absence does not create immunity}.

The framework is two-level. Inference performs two legal functions, \textbf{constitutive} and \textbf{protective}; the protective function operates through three pathways: \textbf{identificatory}, \textbf{attributive} and \textbf{decisional}. \textbf{Composition} is not a fourth pathway but a cross-cutting architectural dimension which, with reach, persistence and reviewability, is what agentic architectures modify. Three concepts support it: the \textbf{inferential threshold}, the \textbf{inferential reach} and the \textbf{inferential chain}, mapped onto the \textbf{chain of imputation}.

Regulation (EU) 2026/1744 left the constitutive criterion untouched and inserted a provision contemplating outputs that influence the inputs of future operations, without supplying any rule of aggregation. The article proposes an interpretive rule, a compositional-effects test identifying the decision unit under Article 22 GDPR together with the allocation of the burden of establishing it, and documentation duties calibrated to inference chains.
\end{abstract}

\vspace{-0.4em}
{\small\noindent\hspace*{0.06\textwidth}\begin{minipage}{0.88\textwidth}
\noindent\textbf{Keywords:} inference; inferred data; agentic AI; AI Act; GDPR; Article 3(1); Article 22; automated decision-making; human oversight; accountability
\end{minipage}}

\newpage
\tableofcontents
\newpage

\section{Introduction}

\subsection{The problem in one example}

An organization deploys an agent to triage inbound applications. The agent retrieves a candidate's submitted documents, queries two internal systems and one external service, summarises what it finds, ranks the file against criteria expressed in natural language, and routes it to one of three queues. No component of this pipeline was designed to determine anything about the candidate's health. Yet the retrieval step surfaces a gap in employment history; the summarisation step characterises that gap; the ranking step weights it; and the routing decision is taken on a file that, by the end of the sequence, contains a derived characterisation that no human wrote, that was present in no source, and that a reviewer reading only the final output would not recognise as an inference at all.

Three questions follow, and EU law answers them from two different directions.

\emph{Is this system regulated as an AI system?} The AI Act answers by asking whether the system infers within the meaning of Article 3(1). The answer may be affirmative for the ranking component and negative for the deterministic router that consumes its output.

\emph{Is the derived characterization protected?} The GDPR answers without asking the first question. It asks whether personal data were processed, whether the derived information relates to an identifiable person, and what that information says about them.

\emph{Who answers for it?} Neither instrument answers directly, because both were drafted with a single system producing a single output, not with a sequence of derivations distributed across components, tools and providers.

\subsection{Two concepts of inference, one word}

The word ``inference'' performs two different jobs in this body of law.

In the AI Act it is \emph{definitional}. Recital 12 identifies the capability to infer as a key characteristic of AI systems and states that this capacity transcends basic data processing by enabling learning, reasoning or modeling. It is the feature that separates the regulated object from conventional software. Poretschkin and Naeven observe that, since the Act does not clearly define what inference is, a grey area persists for certain data-driven systems, and record that, to the best of their knowledge, no prior work had analyzed the inference concept of the AI Act in greater technical detail \cite{poretschkin2026}.

In the GDPR, it is \emph{consequential}. The Regulation does not define inference, does not use the term in its operative provisions, and does not condition protection on any technological qualification. It attaches obligations instead to what an inference \emph{is} (personal data), to what it \emph{concerns} (potentially a special category), and to what it \emph{does} (potentially a decision). The Court of Justice has developed each strand separately, without consolidating them under a single heading.

\subsection{Claim and contribution}

The claim is not that either instrument is mistaken. It is that practitioners and, increasingly, compliance architectures treat the two perimeters as concentric, as though establishing that a system falls outside Article 3(1) said something about its exposure under the GDPR, or as though satisfying the AI Act's requirements for high-risk systems addressed the GDPR's concerns about inference. Neither implication holds.

The contribution is threefold: a \textbf{two-level analytical framework} distinguishing the constitutive from the protective function of inference, and the protective function into three pathways; an analysis of why \textbf{agentic architectures} convert this conceptual mismatch into an operational one, through composition acting across all three pathways at once; and a distinction between the \textbf{inferential chain} and the \textbf{chain of imputation}, with a proposal for mapping the second onto the first.

\subsection{Terminology}

This article develops, on the specific axis of inference, an argument whose action-oriented counterpart is set out in \emph{Agentic AI --- Technical Autonomy, Human Responsibility, and Legal Governance} \cite{fabiano2026}. It follows the distinction developed there between \emph{Agentic AI} and \emph{AI agent(s)}, which are not synonyms \cite[ch.~1]{fabiano2026}. ``Agentic AI'' designates the class of architectures characterized by operational delegation, the capacity to act, not merely to generate, while ``AI agent'' designates a system instance within such an architecture. Where the argument concerns a component rather than a system, this is stated.

The article also treats \emph{agenticity} as a graded property rather than a binary attribute \cite[ch.~2]{fabiano2026}, and treats the AI Act's constitutive criterion as similarly resistant to binary application.

\subsection{Scope limits}

The analysis is confined to EU law. It does not undertake a comparative treatment of regimes that expressly include inferences in the definition of protected information. It is doctrinal rather than empirical: no user study, benchmark or system evaluation is offered. Its treatment of the pending data strand of the Digital Omnibus package is compartmentalized in Section~\ref{sec:legislative} so that the remainder of the argument does not depend on a legislative file whose outcome is unsettled.

\section{Related Work and Positioning}

\subsection{The inferential-analytics literature and its 2019 horizon}

The foundational treatment of inference under EU data protection law remains Wachter and Mittelstadt's \emph{A Right to Reasonable Inferences} \cite{wachter2019}. The article demonstrates that data subjects are granted little control or oversight over how their personal data are used to draw conclusions about them, and proposes a new right addressed to what the authors call high-risk inferences, those that are privacy-invasive or reputation-damaging and have low verifiability by virtue of being predictive or opinion-based. Its intervention is normative: it proposes that the \emph{quality} of conclusions drawn about a person, and not only the procedure by which they are drawn, should be a subject of legal constraint.

Two features bear on the present argument. First, its horizon is pre-2024: it necessarily precedes the AI Act and therefore does not engage with a definitional use of inference. Second, it precedes the decisions in \emph{OT}, \emph{SCHUFA}, \emph{EDPS v SRB} and \emph{Dun \& Bradstreet Austria}, which have since supplied much of the doctrinal material the 2019 proposal treated as absent.

\subsection{Technical readings of Article 3(1)}
\label{sec:technical}

Poretschkin and Naeven \cite{poretschkin2026} supply the most detailed technical analysis of the constitutive threshold to date. Drawing on statistical learning theory, they propose a five-level framework grading the extent to which data shape the input--output mapping of a system: fixed mapping (Level 0); parametric adaptation within a fixed structure (Level 1); structural selection among predefined alternatives (Level 2); data-driven structural construction, explicit or implicit (Levels 3a and 3b); and representational construction, where data determine the representational space itself (Level 4).

Their reading of Article 3(1) with Recital 12 is that inference operates as a structural rather than a purely functional criterion: a system exhibits the capability to infer when the form of the input--output mapping is at least partially shaped by data rather than fully specified ex ante by human developers.

It is important to record that this formulation is an \textbf{interpretive proposal advanced by those authors}, not a restatement of the legislative text. Article 3(1) provides only that the system, for explicit or implicit objectives, infers from the input it receives how to generate outputs. The present article adopts the Poretschkin--Naeven framework as an analytical instrument, and attributes it as such throughout.

Their analysis also surfaces a genuine interpretive difficulty in the Commission's non-binding Guidelines on the definition of an AI system \cite{guidelines2025}: paragraph 42 of those Guidelines excludes systems used to improve mathematical optimisation or to accelerate and approximate traditional, well-established optimisation methods such as linear or logistic regression, on the ground that, while such models have the capacity to infer, they do not transcend basic data processing. As the authors note, this sits awkwardly with Recital 12, which states that the capacity to infer \emph{transcends} basic data processing, and leaves open whether Level 1 systems fall inside or outside the definition.

This literature is technical and does not engage the GDPR.

\subsection{GDPR/AI Act overlap scholarship}
\label{sec:overlap}

A body of work addresses the relationship between the two instruments, with a center of gravity in data governance and institutional interplay rather than in the concept of inference.

Holtz and Ledendal examine the overlaps between the two regimes, analyzing their overall relationship, conceptual similarities and differences, and the provisions of the AI Act that explicitly overlap with the GDPR; their primary focus is AI data governance and a detailed treatment of the quality criteria set out in Article 10 of the AI Act, with a view to facilitating GDPR compliance \cite{holtz2026}. At institutional level, a study prepared for the European Parliament's ITRE Committee assesses overlaps and gaps between the AI Act and other digital instruments including the GDPR, the Data Act and the Cyber Resilience Act, concluding that while each instrument is individually well targeted, their interplay generates significant regulatory complexity \cite{graux2025}.

This literature is directly relevant and is not displaced by the present argument. Its object, however, is the articulation of obligations across the two regimes. It does not take inference itself as the doctrinal hinge, and it does not distinguish the constitutive from the protective function of the term.

\subsection{Agentic AI and EU law}

Nannini et al.\ \cite{nannini2026} provide a systematic regulatory mapping for AI agent providers, integrating draft harmonized standards under Standardization Request M/613, the GPAI Code of Practice, the Cyber Resilience Act standards program under Mandate M/606, and the Digital Omnibus proposals of November 2025. They present a taxonomy of nine agent deployment categories, identify agent-specific challenges in cybersecurity, human oversight, transparency across multi-party action chains and runtime behavioral drift, and conclude that high-risk agentic systems whose behavior drifts untraceably cannot presently satisfy the Act's essential requirements. Their object is compliance architecture, not the concept of inference.

Hacker and Holweg \cite{hacker2026} address the regulatory classification of agents under the AI Act and the legal status of autonomous actions under EU contract law. They argue that the capacity of agents to reason, plan, and execute across disparate external systems requires a shift of oversight toward the orchestration layer, where multi-agent interaction introduces novel risks of misalignment, and propose a staggered task-authorization scheme together with a statutory list of non-delegable legal acts. Their problem is agency and action.

Gardhouse, Oueslati and Kolt \cite{gardhouse2026} likewise examine agents against the AI Act, noting that such systems implicate areas of law ranging from agency and contract to tort and labor.

On the supervisory side, the Spanish data protection authority published on 18 February 2026 a substantial guidance document on agentic artificial intelligence from a data protection perspective \cite{aepd2026}, mapping GDPR obligations, controller and processor roles, transparency, records of processing, data subject rights, automated decision-making, impact assessment, data protection by design, onto agentic architectures, and identifying vulnerabilities arising from environmental interaction, service integration, working and management memory, and autonomy.

\subsection{Inferred groups}

The closest adjacent contribution is Nikiforov \cite{nikiforov2026}, who argues that AI inferences generate a collective redress gap: European digital regulation is increasingly group-aware at the level of risk, while the principal gateways to court-based compensation still run through identifiable individuals or consumers, leaving inferred groups fluid, largely invisible and difficult to translate into identifiable claimants.

That work and this one share a premise: inference produces legally significant objects the existing architecture handles poorly, and diverge on the question asked. Nikiforov asks \emph{who can vindicate} the harms inference produces. This article asks a logically prior question: \emph{what does ``infer'' mean in each of the two instruments, and why does the same word perform different legal work?}

\subsection*{Positioning at a glance}

\begin{table}[htbp]
\centering
\small
\caption*{\textbf{Adjacent contributions and the question each asks}}
\begin{tabularx}{\textwidth}{@{}l l X@{}}
\toprule
\textbf{Contribution} & \textbf{Object} & \textbf{Question asked} \\
\midrule
Wachter and Mittelstadt \cite{wachter2019} & Inferences under the GDPR & Should the \emph{quality} of conclusions drawn about a person be legally constrained? \\
\addlinespace
Poretschkin and Naeven \cite{poretschkin2026} & Article 3(1) AI Act & When does a data-driven system exhibit the capability to infer? \\
\addlinespace
Holtz and Ledendal \cite{holtz2026}; Graux et al. \cite{graux2025} & AI Act / GDPR interplay & How are obligations articulated across the two regimes? \\
\addlinespace
Nannini et al. \cite{nannini2026}; Hacker and Holweg \cite{hacker2026} & Agentic systems & How should agentic \emph{action} be classified and governed? \\
\addlinespace
AEPD \cite{aepd2026} & Agentic systems & Which GDPR obligations map onto agentic architectures? \\
\addlinespace
Nikiforov \cite{nikiforov2026} & Inferred groups & \emph{Who} may vindicate the harms that inference produces? \\
\addlinespace
This article & Inference as a concept & What does ``infer'' mean in each instrument, and why does the same word perform different legal work? \\
\bottomrule
\end{tabularx}
\end{table}

\subsection{The gap}
\label{sec:gap}

To the author's knowledge, the literature has not yet systematically examined the mismatch between the constitutive function of inference under Article 3(1) of the AI Act and the protective function of inference under the GDPR, nor traced its consequences for agentic architectures.

That claim is stated as a claim about the author's knowledge and not as an assertion about the state of the field. It rests on a search of publicly indexed sources and should be tested against a full bibliographic novelty review before submission to a peer-reviewed venue; working papers and non-indexed contributions may have anticipated parts of it.

\section{The Constitutive Function}

\subsection{The text}

Article 3(1) of Regulation (EU) 2024/1689 defines an AI system as a machine-based system designed to operate with varying levels of autonomy, that may exhibit adaptiveness after deployment and that, for explicit or implicit objectives, infers from the input it receives how to generate outputs such as predictions, content, recommendations or decisions that can influence physical or virtual environments.

Recital 12 supplies the reasoning. The definition should rest on characteristics that distinguish AI systems from simpler traditional software or programming approaches and should not cover systems based on rules defined solely by natural persons to automatically execute operations. It then identifies the capability to infer as a key characteristic, describes it as covering both the process of obtaining outputs and the capability to derive models or algorithms from inputs or data, and states that this capacity transcends basic data processing by enabling learning, reasoning, or modeling.

Inference is not the only element of the definition; \emph{machine-based}, \emph{varying levels of autonomy}, \emph{adaptiveness} and \emph{explicit or implicit objectives} all appear. It is, however, a central distinguishing feature, since the remaining elements may equally be satisfied by conventional software.

\subsection{What the criterion does}
\label{sec:whatitdoes}

Three consequences follow from placing inference in the definition rather than in the operative provisions.

\textbf{It is constitutive of material scope.} Where a system does not infer, it is not an AI system, and the obligations that the Act attaches to AI systems do not apply to it \emph{in that capacity}, however consequential its outputs may be. The qualification matters: the AI Act also regulates general-purpose AI models, which are separately defined, and attaches duties to operators in respect of objects other than AI systems. What follows from failing the Article 3(1) threshold is therefore the disapplication of the AI-system obligations to that object, not immunity from the Act as a whole, and, as Section~\ref{sec:pathways} shows, still less immunity from the GDPR.

\textbf{It is binary in application but graded in substance.} The Act asks a yes-or-no question of a phenomenon that, under the Poretschkin--Naeven framework, admits of degrees. Their analysis shows the difficulty concretely: Level 0 is clearly outside and Levels 3 and 4 clearly inside, while the classification of Levels 1 and 2 is unclear, with the Commission Guidelines arguably suggesting that Level 1 is insufficient.

\textbf{It is workflow-dependent, not component-dependent.} One of the most consequential findings in the technical literature is that assessment must consider the entire data-processing workflow rather than isolated components. Poretschkin and Naeven show that a binning procedure which is not itself an AI system, because it produces no predictions, decisions or recommendations, nevertheless contributes to the overall inferential capability of the system in which it is embedded, and that assessing capability by reference only to the final model would yield an incorrect result. They further show that human intervention can \emph{destroy} the capability to infer: where an expert entirely overrules data-derived split points, the data-driven character of the mapping is lost.

\subsection{Why this matters}

The third point carries into the agentic analysis. If the constitutive assessment must take the whole workflow, then in an agentic system the unit of assessment is not the model, not the tool, and not the orchestrator, but the composed pipeline. The AI Act's own logic therefore already points toward composition, while the obligations that follow from qualification are drafted around a system producing outputs, not around a sequence producing intermediate derivations consumed internally.

\subsection{The definition after the 2026 amendment}
\label{sec:amendment}

Regulation (EU) 2026/1744, adopted on 8 July 2026, published in the Official Journal on 24 July 2026 and in force from 27 July 2026, amends Regulation (EU) 2024/1689 together with Regulations (EU) 2018/1139 and (EU) 2023/1230 \cite{reg2026}.

Its amendments to Article 3 of the AI Act are confined to point (14): the definition of \emph{safety component} is replaced, and two new points (14a) and (14b) are inserted defining SMEs and small mid-cap enterprises. \textbf{Article 3(1) is not amended.} The remaining amendments concern the saving clause in Article 2(7), the AI literacy obligation in Article 4, the insertion of a new Article 4a, the prohibitions in Article 5, the high-risk classification mechanics in Article 6, data governance in Article 10 (with the deletion of Article 10(5)), notified bodies, regulatory sandboxes and real-world testing, the powers of the AI Office, and the dates of application in Article 113, with obligations for stand-alone high-risk systems deferred to 2 December 2027 and for AI embedded in Annex I products to 2 August 2028.

The constitutive criterion has therefore passed intact through the first revision of the AI Act. This is a matter of record, not of prediction, and Section~\ref{sec:legislative} draws the consequence.

\section{The Protective Function}
\label{sec:protective}

The GDPR governs inference without naming it. It does so through an entry condition and three pathways, each with its own trigger, its own object and its own line of authority. A fourth element, the explanatory duty, is not a further pathway but a corollary of the third.

\subsection{The entry condition: inference as processing}
\label{sec:processing}

Article 4(2) defines processing expansively. Deriving a new attribute from existing personal data is an operation on those data and therefore a processing activity. This is the condition of entry into the protective function: nothing that follows applies unless it is satisfied.

Two consequences are frequently overlooked in practice, and both require careful statement. The inferential operation must be covered by a lawful basis under Article 6 and must remain within the purposes for which the data were collected or, where it constitutes further processing for another purpose, satisfy the applicable compatibility requirements under Article 6(4) and Article 5(1)(b). It does not follow that every derivation requires a separate and autonomous legal basis: one basis may cover a plurality of operations pursued within the same purpose. What does not follow, equally, is that the basis on which data were collected automatically extends to a derivation pursued for a different end.

Similarly, inferential processing must be adequately reflected in the transparency information provided under Articles 13 and 14 and, where relevant to the description of the processing activity, in the record maintained under Article 30. Those provisions require the description of purposes and categories of processing, not an enumeration of every individual derivation. The error worth naming is not under-documentation of particular inferences but the characterization of derivation as a costless by-product of collection.

\subsection{First pathway: identificatory}
\label{sec:identificatory}

Article 4(1) covers information relating to an identified or identifiable natural person, directly or indirectly. In \emph{Nowak} the Court held that written answers in a professional examination and the examiner's comments on them constitute personal data, establishing that evaluative and opinion-based material, and not only observed fact, falls within the definition \cite{nowak}.

The identificatory dimension has since been sharpened. In \emph{EDPS v SRB} the Court confirmed a relative approach to the concept of personal data in the context of a transfer of pseudonymised information, holding that such data need not be personal for every recipient and that identifiability turns on whether a given party has means reasonably likely to be used to identify the person concerned \cite{edpssrb}; the decision builds on \emph{Breyer} \cite{breyer}. At the same time, the Court held that the controller's transparency obligation is assessed from the controller's own perspective at the time of collection, so that the relative character of the concept does not dissolve the duty to inform.

The relevance here is structural. \emph{Whether an inference reaches a person at all} is itself an inferential question, answered by reference to available means. Identifiability is not a property of data; it is a property of the relation between data and a holder. Note also that this pathway already contains, in Article 4(1) read with Recital 26, a threshold of practical rather than theoretical possibility, a feature to which Section~\ref{sec:attributive} returns.

\subsection{Second pathway: attributive}
\label{sec:attributive}

In \emph{OT} the Grand Chamber held that publication of personal data liable to disclose the sexual orientation of a natural person indirectly constitutes processing of special categories of data within the meaning of Article 9(1) \cite{ot}. The data published, the name of a spouse, cohabitee or partner, were not inherently sensitive. What brought them within Article 9 was the operation of comparison or deduction a reader could perform.

Stated in the terms of the judgment: for Article 9 purposes, the non-sensitive character of the inputs does not preclude the processing from revealing special-category information through combination or deduction.

\emph{OT} does not stand alone. In \emph{Meta Platforms} the Grand Chamber held that the processing of data relating to a user's consultation of websites and applications, and to the entering of information on such platforms, constitutes processing of special categories of data where those data allow information falling within one of the categories listed in Article 9(1) to be revealed \cite{metaplatforms}. Two features of that decision bear on the reading advanced here. The qualification is attached to what the data \emph{allow to be revealed} rather than to the nature of the source; and it is attached to data whose sensitive significance emerges only from combination and from the use to which they are put. Read together, the two judgments describe a line of authority rather than an isolated holding in a peculiar factual setting, a point of some importance, since a structural reading of Article 9 built on a single reference concerning the publication of declarations of interests would be correspondingly easier to confine to its facts.

\subsubsection*{An interpretive proposal: the derivability threshold}

This article advances, as its own interpretive proposal and not as an account of the holding, a wider structural reading: under Article 9 the character of the input does not govern; what governs is the character of what can be derived.

So stated, the proposition would be untenable. If every theoretically possible derivation triggered Article 9, virtually all processing of personal data would fall within it, since almost any dataset can be combined with some hypothetical auxiliary source to yield a protected attribute. The proposition therefore requires a threshold, and the threshold should not be invented.

The identificatory pathway already supplies the model. Article 4(1), read with Recital 26, does not ask whether identification is conceivable but whether there exist \emph{means reasonably likely to be used}; \emph{EDPS v SRB} applies that criterion relatively, by reference to the position of a particular holder. That criterion belongs textually to identifiability and is not transposed here to Article 9 as positive law. It is offered as the model for an analogous threshold of \textbf{concrete and contextual derivability}, proposed in the following terms:

\begin{quote}
Article 9 does not require the protected attribute to have been actually inferred by the controller, nor the controller to have acted with the aim of obtaining it, nor the information revealed to be correct. The provision is engaged where the personal data actually processed, considered in the concrete configuration of their collection, combination, disclosure or use, allow the protected attribute to be revealed through an operation of deduction or cross-referencing that is realistically practicable given the means reasonably likely to be used by the entity carrying out that processing.
\end{quote}

The formulation keeps two planes apart, and the distinction is the point of it. The \emph{object} of the assessment is the concrete processing operation: the data actually processed and the manner in which they were collected, combined, disclosed or used. The \emph{measure} of practicability is entity-relative, in parallel with the identificatory pathway. What does not enter on either plane is the general inferential reach of the architecture: that an entity holds tools and permissions from which a protected attribute could be derived is a property of its configuration, not a quality of any processing operation, and it engages Article 9 only where a particular operation processed and related the data from which the derivation is realistically open.

Three further features follow. The threshold is \emph{contextual}: the same data may satisfy it in one processing environment and not in another. It is \emph{practicable rather than probabilistic}: it does not require quantifying the likelihood that a derivation will occur, only that its performance be realistically open. And it is \emph{indifferent to intention and to accuracy}, which is not a concession but a requirement of the case law.

Both limbs have textual support in the judgments. \emph{Meta Platforms} directs the assessment to whether the data processed allow information within one of the categories of Article 9(1) to be revealed, and states that the prohibition applies irrespective of whether the information revealed by the processing operation in question is correct and of whether the controller acts with the aim of obtaining information of that kind (paragraphs 68 and 69). The reference to \emph{the processing operation in question} is what anchors the object of the assessment; the entity-relative measure of practicability is supplied not by Article 9 but, by analogy, by Article 4(1) read with Recital 26 and by \emph{EDPS v SRB}. The analogy is offered as such: the criterion belongs textually to identifiability and is not transposed to Article 9 as positive law.

Adopting the same order of threshold on both pathways has an additional advantage of coherence: it prevents \emph{inferential reach}, as defined in Section~\ref{sec:crosscutting}, from becoming unbounded at precisely the point where the doctrinal stakes are highest.

Subject to that threshold, the consequence is direct. A protection that attaches to results rather than to sources is, by construction, indifferent to how the derivation was achieved, and therefore indifferent to whether the deriving system satisfies Article 3(1) of the AI Act.

\subsection{Third pathway: decisional}
\label{sec:decisional}

In \emph{SCHUFA} the Court held, in the context of credit scoring, that the establishment of a probability value constitutes automated individual decision-making within the meaning of Article 22(1) where a third party draws strongly on that value in deciding whether to enter into a contract with the data subject \cite{schufa}.

The judgment permits the ``decision'' to be located upstream of the formal act producing the legal consequence, where the inferred value effectively determines that act. That formulation is deliberately narrower than the claim that the Court has generally relocated decisions to the point of inference; what it establishes is a possibility, conditioned on the determinative role of the value. Section~\ref{sec:test} builds on that condition.

\subsection{The corollary of the third pathway: explanation}
\label{sec:explanandum}

In \emph{Dun \& Bradstreet Austria} the Court held that meaningful information about the logic involved, within the meaning of Article 15(1)(h), must describe the procedure and principles actually applied in such a way that the data subject can understand which of their personal data have been used in what way in the automated decision-making at issue; it does not extend to an exhaustive explanation of the algorithm or disclosure of the full algorithm \cite{dunbradstreet}. Where disclosure would undermine a trade secret, the controller must supply the information to the competent supervisory authority or court, which balances the interests at stake.

This is not a fourth pathway. Article 15(1)(h) is triggered by the existence of automated decision-making within the meaning of Article 22(1); the explanatory duty is parasitic on the decisional pathway and has no independent trigger. Its content, however, is what makes the decisional pathway operable in practice: the standard is procedural and individualized, not what the model is, but what was done with this person's data to reach this result.

\subsection{Interim synthesis}

Read together, these decisions supply doctrinal building blocks for a law of inference. They address whether a derivation reaches a person, what its content triggers, when it constitutes a decision, and what must be explained about it.

They do not articulate inference as an autonomous legal category. The Court has had no occasion to do so, because each reference presented a discrete question. The unifying category is proposed here; it is not attributed retrospectively to the Court.

\section{A Two-Level Analytical Framework}
\label{sec:framework}

\subsection{The status of the framework}
\label{sec:status}

What follows is a \textbf{functional typology}, not an ontology. Its categories identify legally distinct questions raised by inference; they are not technical forms of inferential reasoning, and no claim is made about entities, properties or membership conditions in any formal sense.

Three features are declared expressly, because a classification that leaves them implicit invites the objection that it is not one.

\textbf{The criterion of division} is the \emph{legal question} that a derivation raises. It is a single criterion, applied at two levels: at the first, whether inference determines regulatory membership or attracts protective consequence; at the second, which protective consequence.

\textbf{Exhaustiveness} is not claimed. The typology exhausts the questions posed by the two instruments examined, not the questions that inference can raise. Other instruments, sectoral, or those of other jurisdictions, may pose others.

\textbf{Disjunction} is not claimed either. A single derivation may travel more than one pathway simultaneously: an attributive inference about an identifiable person that determines an outcome engages all three at once. The pathways are analytically distinct, not mutually exclusive.

\subsection{First level: two legal functions}
\label{sec:functions}

At the first level, inference performs two functions in EU digital law.

\textbf{Constitutive.} Does the system infer, within the meaning of Article 3(1) AI Act? The question determines regulatory membership. Its object is the system. Section~\ref{sec:whatitdoes} sets out its features: constitutive of material scope, binary in application though graded in substance, and workflow-dependent.

\textbf{Protective.} Does an inference, having been made, attract legal consequence by reason of what it is, what it says, or what it does? The question determines the constraints applicable to a processing operation. Its object is the person affected. It is engaged through the entry condition of Section~\ref{sec:processing} and operates through the three pathways below.

The two functions are the subject of the thesis stated in the title. They are not two species of one genus: they are two different jobs the same phenomenon performs in two instruments, and the whole argument of this article is that they do not track each other.

\subsection{Second level: three protective pathways}
\label{sec:pathways}

Within the protective function, three pathways are distinguished by their trigger and their object.

\textbf{Identificatory} --- Article 4(1) GDPR. Trigger: the derived information reaches an identifiable person, given the means reasonably likely to be used by the entity holding it. Object: the relation between information and holder. Authority: \emph{Breyer}; \emph{EDPS v SRB}.

\textbf{Attributive} --- Article 9(1) GDPR. Trigger: the derived information is liable to reveal a protected attribute, subject to the derivability threshold proposed in Section~\ref{sec:attributive}. Object: the content of the result. Authority: \emph{OT}; \emph{Meta Platforms}.

\textbf{Decisional} --- Article 22 GDPR. Trigger: the derived value determines an outcome producing legal effects or similarly significant effects. Object: the effect on the person. Authority: \emph{SCHUFA}; with the explanatory corollary in \emph{Dun \& Bradstreet Austria}.

\subsection{Composition as a cross-cutting dimension}
\label{sec:crosscutting}

Composition, the configuration in which outputs of step $n$ serve as inputs to step $n+1$, with no external observation intervening, is deliberately \emph{not} presented as a fourth pathway.

The reason is structural. A fourth pathway would have to be coordinated with the other three, distinguished from them by its own trigger and object. Composition has neither. It has no trigger of its own, and it has no object of its own, because it operates \emph{on} the objects of the other three: a composed derivation may reach a person, may reveal a protected attribute, and may determine an outcome, all at once and by virtue of the same sequence. Composition cuts across the pathways rather than standing beside them.

It belongs, instead, to a distinct set: the \textbf{architectural dimensions} along which agentic systems modify inferences that were already legally relevant. Four are identified here.

\textbf{Reach} --- the \emph{inferential reach}: the categories of information a system can reasonably derive, given its available sources, tools, permissions, and configured task space. Reach is an \emph{ex ante} possibility and is the object of ex ante governance (Section~\ref{sec:threemoments}).

\textbf{Composition} --- whether derivations feed forward as inputs to further derivations within a single delegation.

\textbf{Persistence} --- whether derivations survive the step that produced them, in working or management memory.

\textbf{Reviewability} --- whether derivations surface to a position competent to assess them, and whether they can be reconstructed after the fact.

The claim that follows is the one the framework exists to make precise: \textbf{agenticity does not create new legal categories of inference; it modifies the reach, composition, persistence and reviewability of inferences already legally relevant.} An attributive inference produced once, reviewed by a human and discarded is one thing. The same attributive inference produced at step three of a fourteen-step trajectory, persisted in agent memory, and consumed by two subsequent steps without surfacing to any reviewer is another, not because the inference changed its legal character, but because those four properties did.

This is why the two levels of the typology and the architectural dimensions form a matrix rather than a scale. Type of inference and degree of agenticity are independent axes: the first records which legal question a derivation raises, the second the architecture within which it is embedded.

\subsection{The aggregation gap}
\label{sec:aggregation}

The three protective pathways are technologically neutral. Nothing in their text confines them to a single derivation, and each can apply to processing occurring within a sequence. Composition therefore raises no problem of applicability.

What the pathways do not supply is a rule for determining \textbf{when several intermediate derivations should be assessed as one legally relevant unit}. Article 4(1) does not say whether identifiability is assessed at each step or on the state resulting from the sequence. Article 9(1) does not say whether an attribute composed across three steps was ``revealed'' by the processing. Article 22(1) does not say whether a decision is the terminal act or the trajectory that determined it.

This is an \emph{aggregation gap}, not an application gap, and the distinction matters: it cannot be closed by insisting that the provisions apply, because they do. It can only be closed by supplying the missing rule. Section~\ref{sec:test} proposes one for the decisional pathway, where the stakes are highest and the case law most developed.

\subsection{The zone of divergence}
\label{sec:divergence}

Crossing the constitutive function with the three protective pathways yields a zone that compliance practice systematically mishandles. Three cells deserve naming.

\emph{Outside the constitutive function, inside the protective.} A deterministic rule engine that combines two fields to produce a health-related characterization is, on the Commission Guidelines' reading, plausibly outside the AI system definition. Subject to the derivability threshold, it is inside the attributive pathway, and potentially inside the decisional one.

\emph{Inside the constitutive function, outside the protective.} A model trained and operated exclusively on data that are not personal for the operating entity, in the sense confirmed in \emph{EDPS v SRB}, satisfies the constitutive threshold while the identificatory pathway is not engaged for that entity, and the identificatory pathway being the gateway, neither are the others.

\emph{Inside both, on different objects.} The most common case and the most treacherous. The AI Act's obligations attach to the system and its lifecycle; the GDPR's attach to specific processing operations and specific individuals. A system may be compliant as a system while a particular inferential operation within it is not lawful as processing.

\section{Why Agentic Architectures Expose the Non-Coincidence}

\subsection{Chained inference}
\label{sec:chained}

An agentic system does not produce an output; it produces a trajectory. Retrieval, planning, tool invocation, intermediate summarisation, re-planning and execution each transform the state on which the next step operates. Nannini et al.\ characterize agents precisely by the autonomous planning, tool invocation and multi-step action chains this involves \cite{nannini2026}.

Legally, the consequence is that the \emph{unit} on which the three pathways operate, a derivation attributable to a moment, dissolves. Section~\ref{sec:whatitdoes} noted that the constitutive assessment must, under the Poretschkin--Naeven reading, consider the whole workflow. The GDPR contains no express workflow-level test equivalent to that proposal. Its provisions can govern each processing operation occurring within a trajectory; what they do not specify is when a sequence of derivations should be treated as a legally relevant whole.

\subsection{Perimeter fragmentation}
\label{sec:fragmentation}

An agentic system is a composite. Its components may sit at different levels of the Poretschkin--Naeven scale: a language model at Level 4, a retrieval ranker at Level 3, a threshold rule at Level 0. The orchestration layer sequencing them may itself be deterministic.

Where the composed system is assessed as a whole, the highest-level component plausibly determines the classification. It should be stressed that this is an analytical implication of the Poretschkin--Naeven framework and not a rule of positive law; the authors themselves introduce it as a rule adopted tacitly in the course of their analysis. That candor is instructive: even the technical literature most directly concerned with the threshold has no settled rule of composition.

Where components are provided, deployed and assessed by different parties, the workflow-level assessment the constitutive criterion demands may never be performed by anyone. Hacker and Holweg's argument for shifting oversight to the orchestration layer addresses the same structural fact from the direction of accountability \cite{hacker2026}.

Two observations follow. First, fragmentation does not reduce exposure on the protective side: the pathways attach to processing operations regardless of how the system is carved up for AI Act purposes. Second, it is precisely in the fragmented case that the assumption of concentric perimeters is most dangerous, because a provider may reason from a defensible conclusion about its own component to an indefensible conclusion about the composed system's data protection posture.

\subsection{Runtime composition of sensitive attributes}
\label{sec:runtime}

Section~\ref{sec:attributive} argued, subject to the derivability threshold, that the attributive pathway is engaged by what can be derived. Agentic systems generalize the problem in a specific way: they compose at \emph{runtime}, from sources selected dynamically, in combinations not fixed at design time.

The distinction matters because the regulatory imagination has largely been shaped by the training-data case, sensitive data present in a corpus, to be filtered, minimized or removed. Runtime composition is structurally different. There is no corpus to filter. The sensitive attribute exists in no source; it comes into existence within the trajectory, as a function of which sources the agent consulted and in what order.

The AEPD's guidance registers the underlying architectural facts, environmental interaction with internal and external sources, service integration, and the persistence of personal data in both working and management memory, and stresses that controllers must know where personal data sit across the agentic architecture, including in memories and logs \cite{aepd2026}. What remains undertheorised is the legal characterization of an attribute that is neither collected nor stored as such, but composed and then consumed.

It should be stated plainly that this section is the argument's most exposed point. The claim is architecturally plausible and consistent with the vulnerabilities the AEPD identifies, but it is not established by measurement. Section~\ref{sec:limitations} sets out what would be required to establish or qualify it.

\subsection{A legislative acknowledgment of recursion}
\label{sec:recursion}

Regulation (EU) 2026/1744 inserts a new Article 4a into the AI Act governing the processing of special categories of personal data for bias detection and correction. Article 4a(1) restates, for providers of high-risk systems, the conditions previously in Article 10(5), which is deleted, namely that the objective cannot be effectively achieved by processing other data including synthetic or anonymised data; that technical limitations on re-use and state-of-the-art security and privacy-preserving measures including pseudonymisation apply; that access is strictly controlled and documented; that the data are not transmitted or otherwise accessed by other parties; that they are deleted once the bias has been corrected or the retention period ends; and that the record of processing states why the processing was strictly necessary.

Article 4a(2) extends the same basis, subject to the same conditions, to providers and deployers of other AI systems and models and to deployers of high-risk systems, where such processing is strictly necessary in view of biases likely to affect health and safety, negatively affect fundamental rights or lead to prohibited discrimination. The provision expressly adds: \emph{especially where data outputs influence inputs for future operations}.

That clause deserves attention out of proportion to its length. It is, so far as the author is aware, the first place in EU AI legislation where the legislature contemplates in terms the configuration this article calls composition, outputs feeding forward as inputs. Three observations follow.

First, the acknowledgment is confined to a single, narrowly bounded context: an exceptional legal basis for processing special-category data for bias detection. It is not a general rule about recursive derivation.

Second, it confirms that the configuration is legislatively visible. The argument of Section~\ref{sec:aggregation} is therefore not that the phenomenon is unrecognized, but that it is nowhere given a rule of aggregation.

Third, and most consequentially, the provision cuts in a direction opposite to the concern of Section~\ref{sec:runtime}. Article 4a is a permission, subject to safeguards, to process sensitive data in order to \emph{detect and correct} bias. It does not address the case in which sensitive information is \emph{created} by composition in the course of ordinary operation. A provision that licenses looking at protected attributes for corrective purposes is not a provision that governs deriving them incidentally, and Article 4a(2) closes by stating expressly that it creates no obligation to conduct bias detection and correction at all.

\subsection{The aggregation problem under Article 22}
\label{sec:salami}

\emph{SCHUFA} permits the decision to be located at the score where a third party draws strongly on it. Agentic systems raise the inverse difficulty: a trajectory may contain no single derivation that determines an outcome, while the trajectory as a whole determines it completely.

Consider a sequence in which each step narrows the space of available outcomes without fixing any. No individual step produces legal effects or similarly significantly affects the person; the final step is a routing operation on an already-determined file. On a step-wise reading, Article 22(1) is engaged at no point. On a trajectory reading, it plainly is.

Nothing in the current text resolves which reading is correct, because the provision presupposes an identifiable decision. Section~\ref{sec:test} proposes a method for identifying it.

\subsection{Attribution across autonomously invoked tools}
\label{sec:tools}

When an agent invokes a third-party service, the immediate invocation may be selected autonomously by the agent within a delegation previously configured by the controller. It does not follow that the resulting processing occurs without instruction: a processor may act on general, pre-configured instructions, and autonomy at the level of the individual call is compatible with instruction at the level of the delegation.

The question is accordingly narrower and harder: whether the resulting processing remains within the scope of the documented instructions, purposes and role allocation. The AEPD's guidance notes that determination of regulatory responsibilities becomes more complex where an agent acts autonomously, that controller and processor roles require specific analysis where agentic systems access third-party services, and that the controller must design and document data flows, identifying for each system the third parties involved and their role \cite{aepd2026}.

The analytical point for present purposes is that role allocation is a question about the \emph{chain of imputation}, and cannot be answered without first reconstructing the \emph{inferential chain}: asking who was competent in respect of a derivation presupposes an account of which derivations occurred.

\subsection{Explaining a chain}

\emph{Dun \& Bradstreet Austria} requires a description of the procedure and principles actually applied, sufficient for the data subject to understand which of their personal data were used in what way. For a single model this is demanding but tractable.

For a trajectory it is a different problem. The procedure actually applied is the sequence; the principles actually applied are distributed across a planner, a set of tool selections, and possibly several models. The explanation that satisfies the standard is not a description of any component but a reconstruction of the path. The standard is moreover expressly individualized: this person's data, this result, which forecloses generic architectural documentation as a substitute.

This yields a requirement that is simultaneously legal and technical: the trajectory must be reconstructible after the fact for a specific individual. That requirement is not met by system-level documentation, and it is not met by logging that records actions without recording the derivations that motivated them.

\subsection{The inferential chain and the chain of imputation}
\label{sec:twochains}

The central analytical proposal of this article is that two chains must be kept apart.

\textbf{The inferential chain} is descriptive and technical. It comprises the sources consulted, the retrieval performed, the intermediate representations formed, the tool outputs received, the derivations made at each step, and the transformations by which each derivation entered the state consumed by the next. It answers: \emph{how did this conclusion come to exist?} It is the \ emph {realized} counterpart of inferential reach: reach is what could have been derived, the chain is what was.

\textbf{The chain of imputation} is normative and organisational. It comprises the positions, human or institutional, competent to configure the perimeter of the delegation, to constrain the sources and permissions available, to observe execution, to interrupt it, and to review its outcome. It answers: \emph{who was answerable for that coming-into-existence?} The concept is developed at length in relation to agentic \emph{action} elsewhere \cite[ch.~9]{fabiano2026}.

The two are not co-extensive and must not be collapsed. A single position in the chain of imputation may be competent in respect of many links in the inferential chain; conversely, a single inferential step may implicate several positions: the party that granted a tool permission, the party that configured retrieval scope, the party that set an acceptance threshold.

The legally significant object is the \textbf{mapping} between them. Where a link in the inferential chain has no corresponding position in the chain of imputation, there is a \textbf{governance gap}: something was derived that no one was competent to constrain. Where a position exists but the corresponding links cannot be reconstructed, there is an \textbf{evidential gap}: someone was competent, but competence cannot be assessed.

This distinction is where the present analysis departs from a straightforward application of the action-oriented framework. The chain of action and the chain of inference are analytically distinct: action is external and observable, inference is internal and, absent deliberate instrumentation, invisible, while being governed by a common apparatus.

\subsection{Human oversight over a sequence}

Article 14 AI Act requires that high-risk systems be designed to be effectively overseen by natural persons. The provision is drafted around a system whose outputs a person can inspect.

Overseeing a trajectory raises three problems it does not address. The first is temporal: meaningful intervention must occur before an irreversible step, which may be reached before any output is available to inspect. The second is representational: the object of oversight is a sequence, and a sequence must be rendered before it can be reviewed. The third is selective: no one can review every trajectory, so oversight becomes a sampling and escalation design problem rather than an inspection problem.

Nannini et al.'s conclusion that high-risk agentic systems with untraceable behavioral drift cannot currently satisfy the Act's essential requirements bears directly here \cite{nannini2026}: traceability is a precondition of oversight, not an adjunct to it.

\section{The Legislative Moment}
\label{sec:legislative}

The 2026 legislative record is directly relevant, but only if two distinct files are kept apart, a distinction frequently collapsed in commentary.

\subsection{What is now law: the AI strand}

Regulation (EU) 2026/1744 is a regulation amending regulations and applies directly. Its substance, as set out in Section~\ref{sec:amendment}, concerns timing, prohibitions, literacy, classification mechanics, governance, the replaced Article 2(7) and the new Article 4a.

What it did \emph{not} do bears on the argument, though the inference that record supports must be stated with care. The Union legislature revisited the AI Act within two years of adoption, under political pressure and with an explicit simplification mandate, and did not touch the definitional criterion. The constitutive threshold, and with it the interpretive difficulty identified by Poretschkin and Naeven, survived unaltered. Recital 3 of the amending Regulation describes its amendments as targeted and addressed to implementation challenges; the definitional ambiguity is not among the challenges it identifies.

The inference properly available is a narrow one. Non-amendment does not establish that the legislature considered the definitional question and declined to resolve it, nor that it considered the question and regarded it as settled: an argument from legislative silence supports neither conclusion in the absence of preparatory material addressing the point, and it would be a weak foundation for a doctrinal claim if it were relied upon as one. What the record does establish is the state of the text. The criterion stands as drafted, and no legislative resolution of the ambiguity is available to the interpreter. The persistence of the difficulty is demonstrated on the substantive grounds set out in Sections~\ref{sec:whatitdoes} and~\ref{sec:technical}, not by the fact of non-amendment; the legislative record is recorded here as context and not as authority.

The practical consequence is nonetheless the one stated: the problem this article identifies is a post-amendment problem, and not one awaiting legislative resolution.

The amendment also inserted, in Article 4a(2), the recursion clause discussed in Section~\ref{sec:recursion}, and did not accompany it with any rule of aggregation.

\subsection{What is not law: the data strand}

The data strand of the Digital Omnibus package, the proposal amending the GDPR, the ePrivacy Directive, the Data Act and the cybersecurity instruments \cite{com837}, has not been adopted, and at the time of writing the Council has not agreed a negotiating mandate. Successive Presidency compromise texts have circulated \cite{st10677}; the text prepared for the Committee of Permanent Representatives in late June 2026 did not proceed, and the file passed to the incoming Presidency without a Council position.

Because the file is unsettled in both directions, no part of the argument is built on it. Three features of the circulating texts are nonetheless worth recording as \emph{scenarios}, since they bear on the pathways identified in Section~\ref{sec:pathways}.

\textbf{Identificatory pathway.} Proposals to codify an entity-relative test in Article 4(1) would legislate what \emph{EDPS v SRB} established judicially. The effect would be to make the assessment turn more explicitly on the means available to a specific entity, which, in an agentic setting, is a moving target, because an agent's means are a function of the tools it may invoke.

\textbf{Attributive pathway.} Proposals to derogate for the incidental and residual presence of special-category data in the development and technical operation of AI systems and models are addressed, on their face, to training-set contamination. They do not reach the runtime-composition problem described in Section~\ref{sec:runtime}, in which nothing was contaminated and nothing retained: an attribute was composed from clean sources and consumed. The same limitation, as Section~\ref{sec:recursion} showed, affects the enacted Article 4a.

\textbf{Decisional pathway.} Proposals to restructure Article 22 from a prohibition with exceptions into a conditional permission would alter the framing of the pathway without resolving the aggregation gap, which is orthogonal to the prohibition/permission axis: whichever way the provision is framed, it still requires an identifiable decision.

\subsection{A drafting observation}

Circulating recitals and operative provisions in the data strand have used, in parallel passages, both ``used to infer outputs'' and ``used to produce outputs'' to describe the same restriction \cite{st10677}. The two are not equivalent. \emph{Infer} denotes the operation; \emph{produce} denotes the result. A restriction on producing outputs from protected data does not obviously reach the use of those data as a basis for a derivation consumed internally that never surfaces. In an agentic setting, where most derivations are consumed internally, the difference is not stylistic.

\section{Proposals}

\subsection{An interpretive rule, and its textual support}
\label{sec:rule}

\textbf{Constitutive scope and protective reach should be assessed independently, and neither should be inferred from the other.}

Concretely: a determination that a system falls outside Article 3(1) has no probative value for exposure under the protective pathways, and a determination that a high-risk system satisfies Chapter III does not establish the lawfulness of any particular inferential operation it performs.

The amended Article 2(7) AI Act supports this rule directly. As replaced by Article 1(2)(b) of Regulation (EU) 2026/1744, it provides that Union law on the protection of personal data, privacy and the confidentiality of communications applies to personal data processed in connection with the rights and obligations laid down in the AI Act, and that, without prejudice to Articles 4a and 59 of that Regulation, the AI Act shall not affect Regulation (EU) 2016/679, Regulation (EU) 2018/1725, Directive 2002/58/EC or Directive (EU) 2016/680.

The scope of that support should be stated precisely. Article 2(7) establishes \textbf{non-substitution and cumulative application}: the two regimes apply concurrently, and compliance with one does not displace obligations under the other. That is exactly what the interpretive rule requires and is sufficient to establish it.

It does not, by itself, establish that the two perimeters are \emph{non-concentric}. A saving clause of this kind is perfectly compatible with perimeters that coincide; it addresses the relationship between the regimes, not the geometry of their scopes. The non-concentricity thesis rests on the analysis in Sections~\ref{sec:framework} and~\ref{sec:divergence}, not on the saving clause. What the clause does is foreclose the most common practical error, treating AI Act compliance as discharging data protection obligations, and it is offered here for that purpose alone.

The corollary of the rule, stated as the thesis of this article: \emph{inferential capability does not determine legal scope, and its absence does not create immunity.}

\subsection{A compositional-effects test}
\label{sec:test}

The gap identified in Section~\ref{sec:aggregation} is an aggregation gap: the pathways apply to sequences, but supply no rule for when several derivations form one legally relevant unit. The following test is proposed to identify the \textbf{legally relevant decision unit} on the decisional pathway. It does not entail that the unit is always the whole trajectory; frequently it is smaller.

\subsubsection*{Operational definitions}

Three terms are used in a defined sense.

A \textbf{step} is a single operation together with the state transition it produces: an operation whose output enters the state consumed by a subsequent operation. Steps are distinguished by the \emph{function} they perform rather than by type, and a single operation may perform more than one. An \textbf{acquisitional} function introduces into the sequence information subsequently relied upon; an \textbf{inferential} function produces a derivation from information already present; a \textbf{decisional} function narrows or determines the space of available outcomes. A retrieval that surfaces material subsequently relied upon is accordingly a step, notwithstanding that it derives nothing: retrieval frequently determines what can be derived at all, and a unit confined to derivations would exclude the operations that shape their content. Operations whose output enters the state consumed by no subsequent operation are not steps, unless the operation produces the terminal outcome under assessment, though they contribute to the record required by Section~\ref{sec:documentation}.

\textbf{Effective foreclosure} occurs where, following a given step, the remaining steps in the sequence do not realistically alter the outcome, whether because they are ministerial, because they operate within a range already fixed, or because no available configuration of subsequent steps would produce a different result.

A \textbf{meaningful human determination} is an intervention by a natural person who has authority to reach a different outcome and information sufficient to reach it independently of the sequence. Formal confirmation of a proposed result, without that authority and that information, is not such a determination.

The \emph{outcome} by reference to which foreclosure is assessed is not any result the sequence produces, but the result that satisfies the effects threshold of Article 22(1): a decision producing legal effects concerning the data subject or similarly significantly affecting them. Foreclosure is therefore a relation to that outcome and not an independent notion. Where a sequence determines a result that does not meet the effects threshold, there is nothing for the test to identify, and the question is disposed of at the first criterion rather than at the second.

\subsubsection*{The test}

\begin{enumerate}[label=\arabic*.]
\item \textbf{Cumulative determination.} Does the sequence, taken as a whole, determine an outcome producing legal effects or similarly significant effects? The question is asked of the sequence, not of any individual step. This is the primary criterion.
\item \textbf{Effective foreclosure.} If so, at which point was the outcome effectively foreclosed? Foreclosure fixes the forward boundary of the unit: steps subsequent to it do not form part of it, whatever their number. Where foreclosure occurs only at the end, that boundary is the terminal step; where it occurs earlier, the unit ends earlier.
\item \textbf{Composition of the unit, by state dependency.} The unit comprises the foreclosure step and every preceding step whose output entered, directly or through subsequent state transitions, the state on which foreclosure operated. The criterion is one of dependency, not of counterfactual causation: it asks what the foreclosing state was composed of, and is answered by tracing provenance backward through the record rather than by asking what would have happened otherwise. A trajectory may accordingly contain steps that precede foreclosure without forming part of the unit, and where foreclosure occurs early the unit is correspondingly small.
\item \textbf{Presumption where dependencies cannot be distinguished.} Where the controller's records do not permit the dependencies required by the third criterion to be traced, every step preceding foreclosure within the automated sequence identified under the first criterion is presumed to form part of the unit. The presumption is rebuttable by evidence that a step's output did not enter the foreclosing state; but the party unable to distinguish dependencies bears the consequence of that inability.
\item \textbf{Unity of delegation (indicium).} That the sequence proceeds from a single grant of operational authority is evidence that its steps form one unit, and its absence is evidence to the contrary. It is neither necessary nor sufficient: several delegations may contribute to one decisional process, and one delegation may generate several distinct decisions. It is accordingly treated as an indicium and not as a condition.
\end{enumerate}

The third criterion is stated in terms of dependency, and not by extending the unit backward to the earliest step whose contribution to foreclosure was \emph{determinative}, because the latter formulation is not consistent with the first criterion. If determination is cumulative, it does not follow that any individual step was determinative, and a criterion requiring one to be identified reintroduces at the boundary precisely the step-wise reasoning that the first criterion rejects. The difficulty is not merely formal. Where a sequence involves stochastic components, a counterfactual ranking of steps by determinative contribution is not reproducible, and a test that depends on such a ranking is not administrable by a supervisory authority or a court. Dependency avoids both objections: it is a property of the record rather than of a hypothetical, and it is assessed by inspection rather than by reconstruction of alternatives.

The order of the third and fourth criteria matters and is not merely expository. The substantive rule is the rule of dependency; the presumption is an evidential fallback that operates only where the record is insufficient to apply it. Stating the presumption first would make the composition of the unit turn on a default rather than on what the sequence in fact did, and would expose the test to the objection, correctly made against any deliberately over-inclusive reading of Article 22, that it widens the scope of the provision by construction.

Two limits of the dependency criterion should be recorded. The first is architectural. In systems that maintain a single undifferentiated context, every item admitted to that context technically enters the state on which subsequent operations act, and dependency ceases to discriminate: the criterion will return the whole sequence, not because that result was presumed but because the architecture makes co-presence and dependency indistinguishable. That is a substantive result rather than a defect: an architecture that cannot separate what it relied upon from what it merely held has, for these purposes, relied upon all of it, but it should be stated, because it is precisely the class of systems with which this article is concerned. The second limit is evidential and is the subject of the following subsection: dependency can be traced only from a record that records dependencies, and the records currently generated by agentic systems frequently do not.

The structure of the resulting unit is not novel. \emph{SCHUFA} treated as the operative decision a step performed by a different actor, upstream of the act that produced the legal consequence, on the ground that the later act drew strongly upon it. What that judgment resolves for a chain of two nodes, the present test generalizes to sequences in which the determinative contribution is distributed across many steps and no single actor holds the whole. The extension is one of architecture rather than of principle.

\subsubsection*{What the test does not address}

The test identifies the \emph{unit}. It does not answer whether the resulting decision was based \emph{solely} on automated processing within the meaning of Article 22(1). That is a separate question, governed by whether a meaningful human determination intervened, and the two must be kept apart: a trajectory may constitute a single decision unit and nonetheless fall outside Article 22 because a person made a genuine determination within it. Conversely, the presence of a nominal human step does not fragment a unit that is otherwise cumulatively determinative.

\subsubsection*{Who must establish the unit}

The test presupposes that the sequence can be reconstructed: that it is possible to establish which steps entered it, what each introduced, and at what point the outcome was foreclosed. That presupposition is not neutral. The trajectory record is held by the controller. The data subject is structurally unable to know which retrieval surfaced which item, whether a protected attribute was derived along the way, or where foreclosure occurred; as to derivations consumed internally, which by definition never surface, the asymmetry is complete.

This bears directly on the calibration adopted here. No deliberately over-inclusive reading of Article 22 has been proposed, and none should be: an interpreter cannot widen the material scope of a provision merely because under-inclusion appears the graver error, and legality, foreseeability and proportionality all tell against it. But a test that is doctrinally restrained and evidentially unallocated produces, in combination, the very under-protection that an over-inclusive reading would have been designed to prevent, by a less visible route. Restraint on the first plane requires allocation on the second, and the two choices cannot be made independently of one another.

Three provisions support the allocation, and the weight each bears should be stated exactly. Article 5(2) GDPR places on the controller the obligation to be able to demonstrate compliance with the principles of Article 5(1). It does not itself provide that every step is to be treated as forming part of the decision unit where the controller cannot show otherwise; that specific evidential consequence is proposed here. The presumption in the fourth criterion is accordingly advanced as giving procedural effect, in this particular context, to the accountability obligation, not as a consequence already contained in it.

Article 15(1)(h), as construed in \emph{Dun \& Bradstreet Austria}, requires a description of the procedure and principles actually applied, individualized to this person and this result, a standard that cannot be satisfied without a record from which the operative sequence can be identified. Its bearing is therefore direct: an explanation that cannot be given because the sequence was not recorded is not a lesser explanation but no explanation at all.

Articles 12 and 19 of Regulation (EU) 2024/1689, for high-risk systems, require capability for the automatic recording of events over the lifetime of the system and the retention of the logs generated. Their contribution should not be overstated. Those provisions do not require that logs record derivations, the dependencies between states, or the point at which an outcome was foreclosed; they provide part of the technical basis from which reconstruction may be possible, within their own scope, and no more. That limitation is itself instructive: it establishes that compliance with the record-keeping obligations of the AI Act does not, without further design, produce a record capable of supporting the test proposed here, which is precisely the case for the inference-aware record set out in Section~\ref{sec:documentation}.

The consequence is the one stated in Section~\ref{sec:documentation} from the side of documentation, restated here from the side of proof: reconstructability is not an incidental feature of compliance but a condition of the enforceability of Article 22 in distributed automated processing. Where the composition of the unit cannot be established, the presumption operates against the party that controls the record.

\subsubsection*{Three applications}

\begin{table}[htbp]
\centering
\small
\caption*{\textbf{Application of the compositional-effects test}}
\begin{tabularx}{\textwidth}{@{}l X X@{}}
\toprule
\textbf{Configuration} & \textbf{Decision unit} & \textbf{Consequence} \\
\midrule
Foreclosure at the terminal step &
Every step's output is traceable into the state on which foreclosure operated; alternatively, the record does not permit dependencies to be distinguished, and the presumption applies &
The unit is the trajectory. Article 22(1) is engaged, subject to the separate question whether the decision was based solely on automated processing. Article 9 is assessed against the data actually processed and the manner in which they were combined and used across the relevant sequence \\
\addlinespace
Early foreclosure &
The outcome is fixed at an intermediate step; subsequent steps are ministerial or operate within a range already determined &
The unit ends at foreclosure and is smaller than the trajectory. Later steps remain relevant to the record and to the explanation owed under Article 15(1)(h), but not to the composition of the unit \\
\addlinespace
Negative case (i): human determination &
The sequence is composed as above, but a meaningful human determination intervenes within it &
Article 22 is not engaged, for want of the \emph{solely automated} requirement. The unit may still be identified; what fails is the qualification of the decision, not its delimitation \\
\addlinespace
Negative case (ii): effects threshold &
No outcome producing legal effects or similarly significantly affecting the data subject arises &
Article 22 is not engaged, for want of the effects threshold. The question is disposed of at the first criterion, and no unit falls to be composed \\
\addlinespace
Both negative cases &
--- &
The entry condition of Section~\ref{sec:processing} and the identificatory and attributive pathways continue to apply to each processing operation within the sequence \\
\bottomrule
\end{tabularx}
\end{table}

This is an interpretive proposal addressed to supervisory authorities and courts. It requires no legislative amendment and is consistent with the Court's approach of locating the decision where determination actually occurs.

\subsection{Documentation calibrated to inference}
\label{sec:documentation}

The accountability principle in Article 5(2) GDPR requires evidence for all controllers; for high-risk AI systems, the record-keeping obligations under Article 12 AI Act require it additionally and independently. Applied to agentic systems, the following are proposed as the minimum content of an inference-aware record:

\begin{itemize}
\item the \textbf{sources consulted} at each step, including tool identity and the scope of the query;
\item the \textbf{derivations made}, distinguished from the actions taken, an audit trail recording what the agent \emph{did} without recording what it \emph{concluded} cannot support a \emph{Dun \& Bradstreet} explanation;
\item the \textbf{state transitions}, sufficient to establish which derivations entered the input of which subsequent step;
\item the \textbf{constraints in force}, so that what was \emph{not} available to the system is reconstructible alongside what was.
\end{itemize}

The last item matters because inferential reach is defined by constraint. A record showing only what happened cannot distinguish a system that was prevented from consulting a source from one that happened not to.

The governing proposition, stated carefully: \textbf{an unlogged inference may become legally irreconstructible.} This is not a claim that unlogged inferences do not exist or produce no effects. It is a claim about evidence: where the inferential chain cannot be reconstructed, neither the \emph{Dun \& Bradstreet} explanation nor the mapping to the chain of imputation can be performed, and accountability fails for evidential rather than substantive reasons.

\subsection{Limits before, evidence during, attribution after}
\label{sec:threemoments}

The three-moment structure developed for agentic action \cite[chs.~1 and 13]{fabiano2026} transfers to inference, and the transfer is where the two frameworks meet.

\textbf{Before --- the inferential perimeter.} Governance of inference begins with constraint on reach, not with review of output. Purpose stability under Article 5(1)(b) and access minimization under Article 5(1)(c) and Article 25 operate here, applied to the sources and tools an agent may invoke rather than to the data a system may store. The operative proposition: \emph{what is not constrained ex ante remains within the system's potential inferential reach.} This is weaker than a claim that everything unconstrained will be inferred, and correspondingly more defensible; it is also more useful, because it identifies reach --- an ex ante possibility --- rather than actual derivation as the object of ex ante governance.

\textbf{During --- the inferential chain as evidence.} The realized counterpart of reach. Section~\ref{sec:documentation} sets out the content.

\textbf{After --- the mapping to imputation.} The question is not whether the system inferred, but which position was competent in respect of each link, and whether that competence was exercised.

\subsection{Operational instantiation}

The three-moment structure and the mapping between chains can be instantiated in the seven-column matrix proposed in \emph{Agentic AI} \cite[ch.~13 and App.~A]{fabiano2026}, with two modifications: the object of the row becomes the inferential trajectory rather than the delegated action, and the oversight column is assessed against the chain rather than the terminal step.

This is deliberately presented as a specialization of an existing instrument rather than as a second framework. The proliferation of parallel governance apparatuses is itself a compliance risk, and there is no reason for inference to require its own.

\section{Limitations and Future Work}
\label{sec:limitations}

\textbf{Method.} The article is doctrinal. Its claims about how agentic systems behave are drawn from the technical and regulatory literature cited, not from measurement.

\textbf{The empirical status of Section~\ref{sec:runtime}.} This is the argument's most exposed point and is flagged as such in the section itself. The claim that agentic systems compose special-category attributes at runtime from individually innocuous sources is architecturally plausible and consistent with the vulnerabilities the AEPD identifies, but it is not established by measurement. Three questions are open: how often such composition occurs in deployed systems; whether it is detectable from logs of the kind currently produced; and whether composed attributes survive to influence outcomes rather than being discarded within the trajectory. A study addressing them would materially strengthen or qualify the argument, and their absence should temper the weight placed on that section. The derivability threshold proposed in Section~\ref{sec:attributive} mitigates but does not remove the difficulty, since practicability is itself an empirical matter.

\textbf{Jurisdiction.} The analysis is confined to EU law and does not address whether the constitutive/protective distinction has analogs elsewhere.

\textbf{Legislative dependence.} Section~\ref{sec:legislative} depends in part on a file that is unsettled. The remainder is constructed so as not to.

\textbf{Novelty.} The claim in Section~\ref{sec:gap} is a claim about the author's knowledge, resting on a search of publicly indexed sources. A systematic bibliographic review across legal databases and preprint repositories should precede submission to a peer-reviewed venue.

\textbf{Open questions.} The identity of the inferring entity in multi-agent sub-delegation; the evidentiary standard for reconstructing an inferential chain in litigation; whether the aggregation gap can be closed on the identificatory and attributive pathways by tests analogous to that proposed for the decisional one; the treatment of inference in technical standardization under Standardization Request M/613; and whether a graded inferential test can be operationalized within conformity assessment.

\section{Conclusion}

\textbf{Two functions, one word.} The AI Act asks whether a system infers in order to decide whether it is regulated. The GDPR asks what an inference does to a person in order to decide how it is constrained. Neither answers the other's question. The amended Article 2(7) AI Act establishes that the two regimes apply cumulatively and that neither displaces the other; that the scopes they define are also non-concentric is established by the analysis, not by the saving clause.

\textbf{The thesis.} Inferential capability does not determine legal scope, and its absence does not create immunity. A system below the Article 3(1) threshold may travel the identificatory, attributive and decisional pathways in full; a system above it does not thereby discharge any data protection obligation.

\textbf{Three pathways, four dimensions.} Within the protective function, inference is governed along three pathways distinguished by trigger and object. Composition is not a fourth: it operates across all three at once, and belongs with reach, persistence and reviewability among the architectural dimensions that agentic systems modify. Agenticity creates no new legal category of inference; it changes what can be derived, how derivations feed forward, how long they survive, and whether anyone can see them.

\textbf{Two chains.} The inferential chain describes how a conclusion came to exist. The chain of imputation describes who was competent in respect of that coming-into-existence. They are analytically distinct, one technical and descriptive, the other organizational and normative, and governed together. The legally significant object is the mapping between them, and the failures worth naming are the two gaps that mapping exposes: derivations no one was competent to constrain, and competence that cannot be assessed because the derivations cannot be reconstructed.

\textbf{Where responsibility lies.} The Union legislature revisited the AI Act in July 2026, left the constitutive criterion untouched, and inserted, in a narrowly bounded provision on bias detection, an express legislative acknowledgment that outputs may influence the inputs of future operations. It did not supply a rule of aggregation. The data strand remains without a Council mandate. In the interim, the burden falls on supervisory authorities and courts, which have been assembling the doctrinal building blocks for a law of inference without yet articulating inference as an autonomous legal category.

Naming the category is the contribution offered here; agentic architectures are the reason naming it has become urgent.

\appendix


\section{The framework at a glance}

The three tables below summarise the framework developed in Sections~\ref{sec:framework}
and~\ref{sec:crosscutting}. They are a reading aid and not a substitute for the qualifications
set out in the text: in particular, the triggers stated for each pathway are subject to the
threshold discussed in Section~\ref{sec:attributive}, and none of the categories is claimed to be
exhaustive or mutually exclusive.

\begin{center}
\textbf{Level 1 --- Two legal functions of inference}\\[0.5em]
\small
\begin{tabularx}{\textwidth}{@{}l l l X@{}}
\toprule
\textbf{Function} & \textbf{Provision} & \textbf{Object} & \textbf{Question} \\
\midrule
Constitutive & Art.~3(1), Rec.~12 AI Act & The system & Is this within the regulated category? \\
\addlinespace
Protective & Art.~4(2) GDPR (entry) & The person affected & What constraints attach to this operation? \\
\bottomrule
\end{tabularx}
\end{center}

\vspace{0.8em}

\begin{center}
\textbf{Level 2 --- Three protective pathways}\\[0.5em]
\small
\begin{tabularx}{\textwidth}{@{}l l l X@{}}
\toprule
\textbf{Pathway} & \textbf{Provision} & \textbf{Object} & \textbf{Trigger and authority} \\
\midrule
Identificatory & Art.~4(1) GDPR & Data--holder relation & Means reasonably likely to be used; \emph{Breyer}, \emph{EDPS v SRB} \\
\addlinespace
Attributive & Art.~9(1) GDPR & Content of the result & Liability to reveal, subject to the derivability threshold (\S\ref{sec:attributive}); \emph{OT}, \emph{Meta Platforms} \\
\addlinespace
Decisional & Art.~22 GDPR & Effect on the person & Determinative role of the derived value; \emph{SCHUFA}, with \emph{Dun \& Bradstreet Austria} \\
\bottomrule
\end{tabularx}
\end{center}

\vspace{0.8em}

\begin{center}
\textbf{Cross-cutting --- Architectural dimensions modified by agenticity}\\[0.5em]
\small
\begin{tabularx}{\textwidth}{@{}l X@{}}
\toprule
\textbf{Dimension} & \textbf{Agentic stress point} \\
\midrule
Reach & Means are a function of invocable tools, not of a fixed dataset (\S\ref{sec:crosscutting}) \\
\addlinespace
Composition & Runtime derivation across steps on all three pathways at once; no rule of aggregation (\S\ref{sec:aggregation}, \S\ref{sec:salami}) \\
\addlinespace
Persistence & Derivations survive in working and management memory (\S\ref{sec:runtime}) \\
\addlinespace
Reviewability & Derivations consumed internally never surface; perimeter fragmentation (\S\ref{sec:fragmentation}) \\
\bottomrule
\end{tabularx}
\end{center}


\section{Verification record}

This article states the law as at 10 August 2026.

\textbf{Verified.} Regulation (EU) 2026/1744 amends Article 3 of Regulation (EU) 2024/1689 at point (14) only, replacing the definition of \emph{safety component} and inserting new points (14a) and (14b) defining SMEs and small mid-cap enterprises. Article 3(1) is not amended. Verified against the text of the amending Regulation as published in the Official Journal of 24 July 2026 \cite{reg2026}.

\textbf{Verified.} Article 2(7) of Regulation (EU) 2024/1689 is replaced by Article 1(2)(b) of Regulation (EU) 2026/1744 in the terms set out in Section~\ref{sec:rule}; new Article 4a is inserted by Article 1(6); Article 10(5) is deleted by Article 1(9)(b).

\textbf{Verified.} Case C-252/21, \emph{Meta Platforms and Others}, judgment of the Grand Chamber of 4 July 2023, ECLI:EU:C:2023:537, is cited in Section~\ref{sec:attributive} for the proposition that data are processed as special categories where they allow information falling within Article 9(1) to be revealed.

\textbf{Open at the date of writing.} The data strand of the Digital Omnibus package (COM(2025) 837; interinstitutional file 2025/0360 (COD)) is not adopted, and the Council has not agreed a negotiating mandate. Section~\ref{sec:legislative} is written to be replaceable without disturbing the remainder of the article. Its status should be re-checked at the date of submission.

\end{document}